\documentstyle[prb, preprint,dcolumn, epsfig, aps ]{revtex}
\begin{document}

\title{Critical currents and vortex-unbinding transitions in
quench-condensed ultrathin films of Bismuth and Tin}

\author{K. Das Gupta,
Swati S. Soman, G. Sambandamurthy\footnote{Currently at Dept. of Condensed
Matter Physics, Weizmann Institute of Science,
Rehovot, Israel.} and N. Chandrasekhar\footnote{e-mail: chandra@physics.iisc.ernet.in}}
\address{ Department of Physics, Indian Institute of Science,
Bangalore, India}

%\twocolumn[\hsize\textwidth\columnwidth\hsize\csname
%@twocolumnfalse\endcsname

\maketitle
\begin{abstract}
We have investigated the I-V characteristics of strongly disordered ultra-thin films of
{\it Bi} and {\it Sn} produced by quench-condensation. Our results show that both these
sytems can be visualized as strongly disordered arrays of Josephson junctions. The
experimentally observed I-V characteristics of these films is hysteretic, when the
injected current is ramped from zero to critical current and back. These are remarkably
similar to the hysteretic I-V of an underdamped single junction. We show by computer
simulations that hysteresis can persist in a very strongly disordered array. It is also
possible to estimate the individual junction parameters ($R$, $C$ and $I_c$) from the
experimental I-Vs of the film using this model. The films studied are in a regime where
the Josephson-coupling energy is larger than the charging energy. We find that a simple
relation $I_c(T)=I_c(0)(1-(T/T_c)^4)$ describes the temperature dependence of the
critical current quite accurately for films with sheet resistance $\sim$ 500$\Omega$ or
lower. We also find evidence of a vortex-unbinding transition in the I-Vs taken at
temperatures slightly below the mean-field $T_c$.

\end{abstract}
\pacs{}
%]
\section{Introduction}
Ultrathin films produced by quench-condensation under highly reproducible conditions,
with extensive atomic scale disorder have been investigated over the past two decades as
model systems to study the interplay between disorder, interactions and
superconductivity. More than a dozen single and multi-component systems - Al, Bi, Be,
Ga, In, Pb, Sn, Nb, Mo-C, Mo-Ge etc,
~\cite{haviland_89,samband_00,kdg_01,bielejec_00,jaeger_87,orr_84,orr_85,nishida_91,lee_90,ephron_96}
have been shown to exhibit an insulator-superconductor transition as their thickness is
increased or disorder is reduced. The sheet resistance at which the transition takes
place, has been observed to vary from $\approx$ 3k$\Omega$ in Mo-C films~\cite{lee_90}
to $\approx$ 20k$\Omega$ in Al on Ge~\cite{liugold}. The nature of this transition in
the $T \rightarrow 0$ limit has attracted maximum attention~\cite{das_99,phillips}, with
the parameters characterizing the superconducting state having received comparatively
lesser attention. It is also unclear at this point, what the order parameter of this
transition is and what similarities in physics, the phenomenon has with metal-insulator
transition in 2-DEG systems. Most of the reported data on these systems do not show
hysteresis in I-V characteristics, the reason being that in most cases they are not
probed with currents close to the critical currents of these films. Many physically
relevant microstructural parameters can, however, be estimated
from the hysteretic I-Vs.\\

In this paper we
present some experimental results on quench-condensed superconducting
films of Bi and Sn. The choice of the two materials were dictated by
certain experimental considerations, to be discussed below.
Earlier studies had classified
materials like Sn, Pb, Ga as `granular' and Bi (particularly Bi
quenched on a thin Ge-underlayer) as `homogeneous'. The basis of this
classification lay in the rather different resistive transitions to the
superconducting state exhibited by these films.
In Pb, Ga and Sn a metallic phase appears to be sandwiched between the
insulating and superconducting phases. This observation has stimulated
the study of a Bose-metal phase in the $T\rightarrow 0$ limit~\cite{das_99}.
It was first suggested by Ramakrishnan~\cite{tvr_89}, that a
`phase-only' picture was appropriate for the granular films. Destruction
of superconductivity in these films was thought to take place by destruction
of phase coherence (via Josephson coupling) between superconducting grains
embedded in a non-superconducting matrix. In the homogeneous films (Bi has
been taken to be a representative of this class) on the
other hand, destruction of superconductivity was thought to take place
via destruction of the amplitude of the wave function. Though the
resistive transitions (R-T traces) as a function of increasing (decreasing)
thickness (disorder)
for Bi and Sn show qualitative difference, we show using experimentally
obtained I-V characteristics and computer simulations, that both
these systems can be understood as
disordered arrays of Josephson-junctions.\\

Some recent STM studies~\cite{ekinci_98,ekinci_99} on these films
suggest that the structure of the
first monolayer of atoms that stick to the substrate may be rather different
that the subsequent upper layers. Cluster sizes in the range of
\mbox{$\sim$ 30 ${\rm \AA}$} were observed in films nominally 10 monolayers
thick. We discuss the relevance of these to the observed properties of our
films. The structure of these films remains a mystery. Recent STM
data is an indication of what one may expect, but since it was taken under
conditions different from experiments such as those of
Strongin {\it et al.}~\cite{henning_99}, Goldman
{\it et al.}~\cite{haviland_89,jaeger_87,orr_84}, and this work,
comparisons of microstructure should be made with an
element of caution.\\

To summarize, there are significant unresolved problems
regarding the processes that
can lead to formation of such clusters at very low ($<$ 10K) substrate
temperatures. Landau {\it et al}~\cite{danilov_96,parshin_96,landau_97}
pointed out that formation of structures such as those reported by
Valles {\it et al.}~\cite{ekinci_98,ekinci_99},
require that incoming atoms be able to perform $\sim$ 500 hops
before they finally settle to their positions. This order of diffusivity is
much more than what incoming atoms striking cold substrates are believed
to have, although such a possibility cannot be ruled out.
We have not addressed this question, but show that several
properties of the superconducting state in these films may be understood if
we accept the presence of superconducting and non-superconducting regions
in close proximity in the film, a case of microstructure similar to that
reported by Valles {\it et al.}~\cite{ekinci_98,ekinci_99}.
Our model bears a close resemblance to the `percolation-type'
model used by Meir~\cite{meir_99} in describing the M-I transition in 2-DEG, and
has been discussed in our earlier work~\cite{kdg_01}.

\section{Experimental setup}
The experiments were done in a UHV cryostat, custom designed for
{\it in-situ} study of ultrathin films, described in our earlier
publications~\cite{samband_00,kdg_01}.
A turbomolecular pump, backed by a oil-free diaphragm
pump, provided a vacuum better than 10$^{-8}$ Torr. The substrate was
either a-quartz or crystalline Sapphire, mounted with adequate thermal
contact on a copper cold-finger, in contact with a pumped helium bath.
The lowest attainable temperature was 1.8K.
The material ({\it e.g.} Bi. Sn)
was evaporated from a pyrolytic Boron Nitride crucible in a Knudsen-Cell
(effusion cell) of the type used in MBE. The temperature
of the cell was carefully controlled to give a steady deposition
rate between 1-5${\rm \AA}$/minute.
The required temperatures
were 650 $^o$C for Bi and  1150 $^o$C for Sn.\\

Two Hall-bar type masks defined the sample, the actual sample size
was 6mm X 1mm. One of the samples had a pre-deposited Ge underlayer
of $\approx$ 10${\rm \AA}$ thickness. This geometry enables us to
evaluate the claim that Ge underlayers may significnatly improve
the wetting properties of the films, and the possibility that
there may be screening effects due to the presence of an underlayer
of a dielectric constant larger than that of the substrate.
The metal flux reached the sample through carefully
aligned holes in successive cryo-shields cooled by liquid Helium and
Nitrogen. These shields also reduce the heat load on the sample and
provide cryopumping, for better vacuum in the
neighbourhood of the sample.
Electrical contacts to the films were provided through pre-deposited
Platinum contact pads, about 50${\rm \AA}$ thick. Four-probe d.c.
measurements were carried out using a standard high-impedance
current source and a nanovoltmeter or an electrometer.

\section{Observations and discussion}
The R-T traces of a series of Bi films exhibiting the I-S
transition is shown in fig. ~\ref{snfilms}.
Fig.~\ref{bi_iv} shows a typical set of observed I-V
characteristics for a film in the superconducting regime.
We can read off the critical and retrapping currents for the whole film.
We claim here (and demonstrate later) that this ratio would be nearly same as
that of a single junction. For a single junction we have the exact
relation between the Josephson coupling energy, charging energy,
normal state resistance and the Stewart-McCumber junction
parameter $\beta$
\begin{equation}
E_J/E_c = (2/\pi^2)(R_Q/R_N)^2\beta
\end{equation}
where $R_Q = h/4e^2$.
With the approximations stated above, for a film
with $R_\Box \approx$ 500$\Omega$ and $\beta \approx$ 2, we
get $E_J/E_c\approx$ 70. This places the films for which we
find a simple empirical relation between the critical current
and temperature, well into the regime $E_J >> E_c$.\\

The I-V characteristics of a single Josephson-junction, in the
framework of the RCSJ
model, is well known and
is discussed in several texts~\cite{tinkham_book}.
If the transfer of a Cooper-pair across
the junction causes negligible change in the phase-difference ($\phi$)
across the junction, then $\phi$ can be viewed as a continuous variable
and the I-V characteristics (neglecting the effect of thermal noise)
can be deduced from eq. ~\ref{jjeq}.
\begin{equation}
\beta\stackrel{..}{\phi} + \stackrel{.}{\phi} + \sin\phi = i
\label{jjeq}
\end{equation}
The long time average $\langle\stackrel{.}{\phi}\rangle$ gives the
observed voltage drop. $i$ is the normalized current, $I/I_c$, through the
junction.  $\beta$ is related to the microscopic
parameters of the junction as $\beta = (2eI_cR_N/\hbar)R_NC$ and also fixes
the ratio of the retrapping to critical currents ($I_r/I_c$).
The larger the value of $\beta$, smaller is this ratio and wider the
hysteresis.

We first investigate, by computer simulations, whether the hysteresis
would persist in a 2-D RCSJ array even if all the junction parameters
are allowed to have a large distribution of values ({\it i.e.} in
presence of strong disorder). Also we need to know whether the observed
I-V curve would be significantly altered by the particular type of
distribution ({\it e.g.} square, Gaussian, log-normal). We address these
questions in the section on computer simulation, the details of which are
presented later.

An important fact about a 2-D network of resistances,
is the following: not too close to the percolation threshold,  the
observed resistance, measured between two edges (see fig.~\ref{jjarray_sch}),
is close to the average value of all the resistances.
In a 1-d chain it is obviously the sum of all the resistances, in a
3-d lattice the measured resistance is much lower than the average.
These can be verified by simple numerical calculations.
In 2-D  even when the width of the distribution is
more than 90\% of the mean value, the measured
normal state sheet resistance ($R_\Box$) of the film does not
differ from the mean of all the resistances by more than 10\%.
We have verified by direct numerical calculation that this result
holds good for array sizes down to a 10$\times$10 array.
If there are $N$
junctions in parallel, then the observed critical current of the
array would be approximately $N$ times the average critical
current of one junction. This allows us to infer some important facts.
This also demonstrates our claim above, that the ratio of the critical
current to the retrapping current for an array is the same as that for
a single junction in the array. In what follows, this will be
further substantiated.

\subsection{Hysteresis in a disordered array}
If identical junctions were laid out on an array, the I-V
characteristics would be indistinguishable from that of a
single junction in absence of a magnetic field. However when
there is a large spread in the values of the junction
parameters, it is not a-priori obvious what their behaviour will be.
This is because different junctions with different critical currents
may undergo transitions at different currents, and for a large
array the sharp transition might be broken into many steps
and then rounded off by finite temperature effects. Our main result
in this regard is that, even in a small array (16$\times$16) disorder
does not completely destroy the hysteresis- till about 50\% disorder
(ratio of width to mean value in a square distribution) the shape of
the I-V loop does not change appreciably. We have shown the
results (see fig.~\ref{simulated_iv}),
using a square distribution of disorder here, though simulations
using a log-normal distribution (results not shown) also lead to
similar conclusion.

\subsubsection{Driving equations and the algorithm}
In an array the
algebraic sum of all the currents meeting at any node must be zero -
this is required by Kirchoff's current law. As
shown in fig.~\ref{jjarray_sch}, the current is fed uniformly through one
edge of the array and extracted through the opposite. In the direction
perpendicular to that of current injection, we generally use periodic
boundary conditions, this is equivalent to joining the remaining two
free edges of the array together. From the current conservation equations
we get a total of $N^2$ coupled second order differential equations.
The model we are considering neglects the self inductance of the array.
Consider a single node, (not on any of the edges)
at which the phase at some instant is $\phi$. In
the four neighbouring nodes the phases are $\phi_u$, $\phi_l$,
$\phi_b$, $\phi_r$ - where the subscripts denote up, left, bottom and
right respectively. Every bond will have its characteristic $R$, $C$ and
$I_c$ value - since the array is disordered. We use $R_0$, $C_0$ and
$I_{C0}$ to denote the average values. With a little algebra we can show
 that the contribution of any link to a node is:\\
 \centerline{$i = c\beta_0\stackrel{..}{\phi} + r\stackrel{.}{\phi} +
   i\sin\phi$},
  where, for a completely ordered array $r=1$, $c=1$ for all bonds.
  Adding all the current contributions to a node we get

  \begin{eqnarray}
  \beta_0[(c_l + c_u + c_r + c_b)\dot{p} & - & c_l\dot{p_l} -
  c_u\dot{p_u} - c_r\dot{p_r} - c_b\dot{p_b}] + \nonumber \\
   (1/r_l + 1/r_u + 1/r_r + 1/r_b)p &-&
    (p_l/r_l + p_u/r_u + p_r/r_r + p_b/r_b) + \nonumber\\
   (i_ls_l + i_us_u &+& i_rs_r + i_bs_b) = 0
  \end{eqnarray}

  \noindent  where $\dot{\phi}_{l,u,r,b} = p_{l,u,r,b}$ and
  $\sin(\phi - \phi_{l,u,r,b}) = s_{l,u,r,b}$. For the nodes on the boundary, there
  will be three contributions from the three nearest neighbours,
  the fourth contribution
  will be $\pm i$, depending on whether current is injected/extracted from the node.
  Thus we get a total of $N^2$ coupled second order differential equations, where
  ($p,\phi$) form a total of 2$N^2$
  variables to be updated at each step. We can visualize
  the whole set of equations, in a matrix form, as:
  \begin{eqnarray}
  {\mathbf C\dot{P} =  RP} & {\mathbf + } & {\mathbf IS = D} \nonumber\\
  {\mathbf \dot{\Phi} } & {\mathbf = } & {\mathbf P}
  \end{eqnarray}
 where ${\mathbf P}$, ${\mathbf D}$, ${\mathbf S}$
 and ${\mathbf \Phi}$ are column vectors of length $N^2$ each,
 ${\mathbf C}$, ${\mathbf R}$ and ${\mathbf I}$ are $N^2{\times}N^2$ matrices.
 These matrices do not change with time. However  ${\mathbf C}$ is a singular
 matrix, irrespective of whether the array is regular/disordered.
 This singularity implies that all the variables in the problem
 are not independent.  This introduces an extra complication in the problem.
 In mathematical literature such systems of
 equations are called ``differential algebraic equations'' (DAE). They occur
 frequently in lattice related problems. \\

 Physically it is not difficult to
 trace the equation of constraint here. We have written equations for the phase
 of each junction, whereas only the phase differences are of consequence. We can
 add an arbitrary number to each phase ensuring\\
 \centerline{${\displaystyle \sum_{all\ \  nodes}} \phi_i = 0$} \\
 This is achieved by setting all the numbers in any row (say the last) of the
 matrix ${\mathbf C}$ to be equal to one and the corresponding entry in the
 column vector ${\mathbf D}$ to be zero. The modified ${\mathbf C}$ is no longer
 singular and can be inverted. The set of equations is solved by using a
 fourth-order Runge-Kutta method with variable
 time-stepping~\cite{numrecipe_book}.\\

 For each current we allow the system to evolve till $\tau = 2500$, the first 500 time
 units are regarded as stabilization time and discarded. The voltage across the array
 is then averaged between 500$< \tau <$ 2500. The value of the external current
 is then increased by a small amount and the above-mentioned cycle repeated, till the
 injected current reaches the desired maximum value. After that it is decremented in
 identical steps. The recorded values of the voltage vs drive-current is the response
 of the array to a cyclic external current. The $I-V$ curve obtained this way
 is hysteretic for $\beta_0 > $1, as expected. \\

 In this algorithm every step requires a large ($N^2{\times}N^2$)
 matrix multiplication.
 This requires $N^4$ multiplications for each update of the variables. Assuming that
 in each step we change the current fed to the array by $I_C/N$, where $I_C$ is the
 average critical current of each bond, the total complexity of the simulation
 increases as $\sim N^5$.

 \subsubsection {A note on the ``fast'' algorithm for {\it regular} arrays}
For ordered arrays the computation can be made much faster, exploiting the special
form the matrix ${\mathbf C}$ takes in such a case. The matrix ${\mathbf C}$ is
then a ``connectivity matrix'', the $(\vec{r},\vec{r^{\prime}})$ element of the
matrix will be -1, if the sites $\vec{r}$ and $\vec{r^{\prime}}$ are connected by
a bond, the diagonal element will give the co-ordination number of the site.
Himbergen {\it et al} ~\cite{eikmans_90} noted
that this particular form of the
matrix allows the multiplication to be carried out in ${\sim} N{\ln}N$ steps.
Because the eigenvectors of this matrix are of the form $\exp(i\vec{k}.\vec{r})$,
the multiplication with the inverse of the matrix can be viewed as two discrete
Fourier transforms, amenable to ``fast Fourier transform'' techniques.
The technique was improved by Dominguez {\it et. al.}~\cite{dominguez_91}
and applied to several
array geometries soon after~\cite{datta_96}.
Unfortunately, this fast algorithm requires that the capacitance of all the bonds
be same, even though disorder in $R$ and $I_C$ can be handled. We however need to
see the effect of disorder in all the bond parameters. Consequently we had to
use  straightforward matrix multiplication - i.e. the ``slow'' technique.

\subsection{Critical currents of the films}
We next investigate the temperature dependence of the experimentally
measured critical currents. Here we plot the normalized critical current
against the reduced temperature.
Fig.~\ref{bi_sn_ic} shows data from several Bi films, gathered from
different runs and on different substrates. All the points
appear to collapse on a simple power law curve, given by
\begin{equation}
I_c(T)/I_c(0) = 1 - (T/T_c)^4
\end{equation}
 A similar behaviour of Sn films is also shown in fig.~\ref{bi_sn_ic}.
The critical current of a weak-link is related to the
superconducting gap ($\Delta$) by the
well known Ambegaokar-Baratoff relation
\begin{equation}
I_cR_N = (\pi\Delta(T)/2e)\tanh \Delta(T)/(2k_BT)
\label{AB-formula}
\end{equation}
As $T\rightarrow T_c$, $\Delta\rightarrow 0$, we can expand the tanh term to show that
$I_c \sim \Delta^2$. This suggests that for these films $\Delta$ vanishes as $\sqrt{1 -
(T/T_c)^4}$. Near $T_c$, we have  $\sqrt{1-(T/T_c)^4}\approx 2\sqrt{1-(T/T_c)}$, which
is consistent with the behaviour of a BCS gap, as far as the leading power is concerned.
However inserting the BCS-result, ({\it i.e.} $\Delta(T)/\Delta(0) \approx 1.74\sqrt{1-
T/T_c}$) in the Ambegaokar-Baratoff relation leads to the well-known prediction that
near $T_c$, the slope ($I_cR_N/(T-T_c)$) should be 635${\mu}V/K$. For all the Bi and Sn
films studied by us, we found this slope to be 960$\pm$20${\mu}V/K$.

%This is not the behaviour of a BCS gap near $T_c$, which varies as $\sqrt{1-(T/T_c)}$.

It is interesting to note here that a similar behaviour of $\Delta(T)$ ({\it i.e.}
$\sim\sqrt{1-(T/T_c)^4}$) over the entire range of temperature has been observed in
fabricated Josephson junction arrays ~\cite{delsing_94}. In such fabricated arrays, one
naturally expects the spread of the junction parameters to be quite narrow, and in this
sense the disorder is considerably less than that for a random array, such as the quench
condensed films we study in this work.

Though the data sets are restricted to  $T/T_c >  0.3$, in all the experiments, the
flattening of the curves allows us to make an extrapolation of the critical current to
the $T\rightarrow 0$ limit. This is important, since $I_c(0)$ is very simply related to
the superconducting gap by eq. ~\ref{AB-formula}. This relation can be refined further,
to account for the presence of disorder, as has been done by Kulik and
Omel'yanchuk~\cite{kulik_75}. In fact, disordered films may show a much better match to
the Kulik and Omel'yanchuk form, rather than the Ambegaokar-Baratoff form. Since the
Kulik and Omel'yanchuk form comprises recursive functions, no simple expression exists
for the dependence of the gap/critical current on reduced temperature~\cite{kulik_75}.
However, the corrections for the presence of disorder to the Ambegaokar-Baratoff
relation, near $T=0$ is within a factor of two, for both the dirty and the clean limits.
This increase for the dirty limit may partly explain the reason for the steeper
dependence of $i$ on $t$; observed in this work. Very close to $T_c$ it coincides with
eq.~\ref{AB-formula}.

As an interesting aside, we mention that
Dynes {\it et al} have shown  in tunneling experiments on
quenched Sn and Pb films ~\cite{valles_89},
that the ratio $2\Delta(0)/k_BT_c$ remains
close to 3.5 (the BCS value) for sheet resistances upto at least 4k$\Omega$.
The agreement was better for Sn than Pb, this may be expected as Pb is
a ``strong-coupling'' superconductor.

\subsection{Estimate of the number of junctions in a film}

Using the model of  disordered array as the background and the extrapolated
values of the critical current at $T=0$, we show that an estimate of the
number of junctions in the film can be made. The estimate shows that
not all grain boundaries  may be acting  as  junctions/weak-links.
It also supports the
possibility that the first layer of atoms that stick to the substrate
may have a significantly different structure~\cite{ekinci_98,ekinci_99}
than the subsequent upper
layers. In such cases a slightly more uniform lower layer may offset the
phase-breaking effect of a considerable number of grain-boundaries.\\

If superconducting behaviour of each grain follows approximately the
BCS model, then we should have the zero temperature gap
\mbox{$\Delta(0) = 1.76k_BT_c$}. Such an assumption is certainly valid
in the vicinity of $T_c$, deviations from this being important only
at lower temperatures, as discussed above.  The average critical current of
each junction is $I_c/N$, where $I_c$ and $T_c$ are experimentally
measured. The number of junctions acting in parallel
should then be given by
\begin{equation}
 I_cR_N/N = \pi\Delta(0)/2e = \pi(1.76k_BT_c)/2e
 \label{N_junc}
 \end{equation}
The total number of junctions (over 1mm$\times$1mm) is then
approximately $N^2$. Using the critical current data shown and
their $T_c$ we find the following for the Bi films:

%\begin{table}
%\caption{The estimates of the number of junctions for Bi films.
%This estimate is for a 1mm$\times$1mm area.}

\begin{center}
\begin{tabular}{|c|c|c|c|c|c|}
\tableline dataset & thickness           &  $T_c$(K)  & $R_N$($\Omega$)    & $N$ &
$4I_c(0)R_N/T_c$\\\hline
\#36 &  53${\rm \AA}$Bi    & 4.32 &  421  &  942 & 953\\\hline
\#25 &  65${\rm \AA}$Bi/Ge  &   4.5     &  236  &  651 & 954\\\cline{2-6}
     & 65${\rm\AA}$Bi & 4.29    &  252 &  730 & 955\\\hline
\#35&85${\rm\AA}$Bi/Ge & 5.23 & 135&  473 & 954\\\hline \tableline
\end{tabular}
\end{center}
%\end{table}

We find that there are $\sim$ 10$^6$ junctions/mm$^2$ in a
$\approx$60${\rm \AA}$ film.
As expected the number of junctions reduce when the film thickness is increased.
This is expected as many of the gaps/voids between grains may be filling up
as more material is deposited, reducing the number of junctions.
Based on this we obtain the following values for $\omega$, $R$ and $C$ for the
85${\rm \AA}$ film at $T$=0:\\
$\beta = 4.5$, $\omega = 3.63\times10^{12} rad/sec$,
$I_c$/bond = $8.9\mu A$, $C_{bond} = 9\times 10^{-15}$Farad, \\
which are reasonable values.

We find
that at low temperatures $I_c$ tends to a constant value and so does $I_r$. This
implies that the ratio $I_r/I_c$ and hence $\beta$ also varies very little at
low temperatures, which is to be expected.
Near $T_c$ the behaviour is dominated by the variation
of the critical current of the bond. The value of the capacitance and
resistance (normal state)
do not vary much with temperature. We also find that for this value of thickness,
the film is in a regime where  charging energy/``coulomb blockade'' is
negligible compared to the Josephson-coupling energy at low temperatures.
The value of capacitance will however reduce drastically if the thickness of
the film is lesser. In fact a recent optical
frequency measurement ~\cite{henning_99}
of the intergrain capacitance, in much thinner Pb films, reported a value
of $\sim$ $2\times 10^{-19}$ Farad,
which can be compared with the values we have inferred above.
In such cases single electron tunneling
effects may be expected to play a very important role in transport processes.
It is interesting to compare the values we have
estimated with typical values of ``fabricated'' regular Josephson-junction arrays.
From the published literature we pick one work ~\cite{delsing_94}
we have already cited earlier.
We find that the arrays used had a typical junction capacitance
of 1-3$\times$10$^{-15}$Farad, junction area of $\sim$ 1$\mu$m$^2$, junction
resistance of 4-150k$\Omega$. These values, particularly those of junction-capacitance
and ``unit cell''-area
that we have estimated, are of the same order. However the value of Josephson
coupling energy (and hence $I_c$/bond) reported by them are much less,
compared to deposited films.  Thus
the screening effects of the supercurrents flowing in the films may be expected
to be much greater than in fabricated arrays.
The current density would surely have a lot of spatial variation- if however
we deliberately neglect this aspect and calculate a  supercurrent
density a film (taking the nominal thickness to be the average
thickness) can support before going normal, for the 85${\rm\AA}$ film we
get a number $\sim$ 10$^5$Amp/cm$^2$, which can be compared with typical values
reported for disordered films of the copper oxide superconductors.

\subsection{A possible Kosterlitz-Thouless (K-T) transition in presence of strong
disorder}

In an ordered array of Josephson junctions, at finite temperatures
vortex-antivortex pairs are generated spontaneously. These vortices may
be visualized as a circulating pattern of the ``phase-variable" in
neighbouring islands, Several characteristics of these vortices have been
studied in superfluids and arrays. One of the atomic scale disordered systems that has
been studied with respect to the K-T transition is the quenched
Hg-Xe mixture ~\cite{kadin_83}.
In this section we investigate whether some of the observed characteristics
of the I-V curves of disordered Sn and Bi films can be attributed to a
Kosterlitz-Thouless type transition.\\

Fig.~\ref{bi_kt_40a}  shows a set of  I-V curves taken at various
temperatures. All the curves show a clear critical current and retrapping
current- and a  transition to the normal state resistance of 1210$\Omega$
and 942$\Omega$ for this  particular film.
For a few temperatures just below $T_c$,
the ``superconducting state'' shows dissipation. The resistance remains
constant over 3-4 decades of current and hence there is no
self heating effect.
It is tempting to identify the appearance of a ohmic dissipative state
with the unbinding of vortices - the ``Kosterlitz-Thouless'' transition.
However we need to be cautious with such an identification. One of the
signatures of K-T transition is that, below $T_{KT}$ there is a regime where
the voltage increases as the cube of the current. The reason for this is that
below $T_{KT}$ all the free vortices are generated by the current itself. The
current exerts oppositely directed  ``Lorentz force'' on the vortex and the
antivortex - and hence tends to create free vortices by breaking the pairs. The
number of vortices just below $T_{KT}$ increases as $I^2$ and hence the voltage
rises approximately as $I^3$. Even though we may see such linear regions in
a log plot, we do not always find this predicted cubic dependence in that region
of temperature. Experiments on proximity coupled arrays ~\cite{resnick_81}
have shown both - the linear regime in a log plot and cubic dependence of
voltage on current.
In our case we find that the 70${\rm \AA}$ Bi film data shows this feature - see
fig.~\ref{bi_kt_70a} (upper panel), whereas for many other thicknesses
the {\it I-V} curves below the possible $T_{KT}$ do not
have any region where a $V\propto I^3$ dependence is obvious. Any such behaviour
should have shown up clearly in a log-log plot.
The 70${\rm \AA}$ film also shows another characteristic signature of $K-T$
transition (lower panel). Above $T_{KT}$ the resistance rises with temperature as\\
\centerline{ $R(T) = R_0 \exp[-\alpha/\sqrt{T-T_{KT}}]$}.
The ``best fit'' is shown in fig.~\ref{bi_kt_70a}. We  have a mean-field
$T_c$ = 3.55K and $T_{KT}$ = 3.01K for this film.
With increasing thickness, the resistive transition
becomes steeper and the range of temperature over which the K-T type
behaviour may be seen, also narrows down. For a 100${\rm\AA}$ Bi film
($R_\Box$ = 106$\Omega$, data not shown) we found that the region
narrows down to less than 100mK. This is to be expected since the thicker
films approach 3-D behaviour and the K-T transition is restricted to
2-D systems. An important aspect that remains unaddressed is the robustness
of the K-T transition to disorder. However, since vortices are macroscopic
objects, which average over large areas of the films - compared to
microscopic atomic scale disorder, the K-T transition may be expected to
remain quite robust in presence of disorder.

\section{Conclusion}

We have shown that several properties of quench condensed films of Sn and Bi can be
understood by visualizing them as strongly disordered arrays of Josephson junctions.
Superconductor-insulator transition in such arrays are well known  and predicted to
occur around $E_j \approx  E_c$ ~\cite{geerligs_89,zwerger_book}. Refering to
eq.~\ref{jjeq} we can see that it should occur in the vicinity of $R_N \approx R_Q$ but
not necessarily exactly at $R_N = R_Q$. A simple power-law behaviour of the critical
current of these films (with $R_\Box < $ 500$\Omega$) is found, and the observed I-V
characteristics indicate a possible vortex-unbinding transition in these films. The
picture of Josephson coupling between superconducting patches in the film, implies that
there is a strong variation of carrier density in the film itself. This bears a strong
resemblance to the percolation type description of 2-DEG ~\cite{meir_99} that has been
successfully used to describe some aspects of the observed metal-insulator transition in
Si-MOSFETS.
 To some extent computer simulations can reproduce the
behaviour of the actual films. The difficulties in doing numerical work
on strongly disordered systems are well known.
Though a renormalization group analysis of the K-T transition has been
done ~\cite{kadin_83}, the effect
of strong screening ({\it i.e.} large self-inductance) in a disordered
array remain to be investigated.

\section{Acknowledgements}
This work was supported by DST and UGC, Government of India. KDG thanks
CSIR, Govt. of India for a research fellowship. We acknowledge discussions with Professors
A. M. Goldman and C. J. Adkins.

%\newpage
%\centerline{\bf FIGURE CAPTIONS}

\begin{figure}
\caption{\label{snfilms}Resistance-Temperature (R-T) curves of a set of quench-condensed Sn films showing the insulator-superconductor transition.}
\end{figure}

\begin{figure}
\caption{\label{bi_iv}Current-Voltage (I-V) characteristics of Bi films on Sapphire. The
observed characteristics are similar for Bi and Sn on crystalline as well
as amorphous substrates.}
\end{figure}

\begin{figure}
\caption{\label{jjarray_sch}Schematic of a $5\times5$ JJ-array. each bond is
a parallel combination of three elements, as shown. The current is injected
and extracted as shown, in the other direction we use periodic boundary
condition.}
\end{figure}

\begin{figure}
\caption{\label{simulated_iv}Computer simulated I-V curves of a
$16{\times}16$ array of Josephson junctions, on a square lattice.
The bond parameters ($R$,$C$,$I_c$) are chosen from a square
distribution- `w' denotes the width of the distribution as a
fraction of the mean value of the parameters. The mean values are
used to calculate $\beta$. Use of a `log-normal' distribution also
leads to similar results. }
\end{figure}

\begin{figure}
\caption{\label{bi_sn_ic}Critical currents of a set of Bi and Sn films
of low sheet resistance, in the regime $E_J >> E_c$. Both show a similar
power law behaviour over the entire temperature range.}
\end{figure}

\begin{figure}
\caption{\label{bi_kt_40a}Data from a 40${\rm\AA}$ film on bare quartz (upper panel)
and with 10${\rm\AA}$ Ge-underlayer (lower panel). Although the I-V's are hysteretic, the
hysteresis is not shown for purposes of clarity. In both cases, a few I-V curves
below $T_c$ have a linear part. Data taken at lower temperatures do
not show a power law behaviour (linear region in a log-log plot).
}
\end{figure}

\begin{figure}
\caption{\label{bi_kt_70a}Data from a 70${\rm\AA}$ Bi film shows power law region in log
plot (upper panel) as well as the predicted dependence of R on T (lower panel). Both of
these features together identify a K-T transition. Hysteresis is not shown for the sake
of clarity.}
\end{figure}

%\vspace*{10cm}

%\newpage

%\epsfig{file=fig1.eps, width=15cm, clip=}

%\clearpage

%\epsfig{file=fig2.eps, width=15cm, clip=}

%\clearpage

%\epsfig{file=fig3.eps, height=21cm, clip=}

%\clearpage

%\epsfig{file=fig4.eps, width=15cm, clip=}

%\clearpage

%\epsfig{file=fig5.eps, height=21cm, clip=}

%\clearpage

%\epsfig{file=fig6.eps, height=21cm, clip=}

%\clearpage

%\epsfig{file=fig7.eps, width= 15cm, clip=}

%\clearpage

\end{document}